\documentclass[
  journal=pasa,
  manuscript=research-paper, %% or "review"
  year=2022,%year=2020,
  ]%volume=37,]
  {cup-journal}

\usepackage{microtype,siunitx,booktabs}
\usepackage{amsmath} 
\usepackage{caption}
\usepackage{subcaption}
\usepackage{overpic}

\usepackage[colorlinks=true, allcolors=blue]{hyperref}

\usepackage{xcolor}

\title{Swimming by spinning: spinning-top type rotations regularize sperm swimming into persistently symmetric paths in 3D}

\author{Xiaomeng Ren}
\affiliation{Department of Engineering Mathematics \& Bristol Robotics Laboratory, University of Bristol, BS8 1UB Bristol, UK}

\author{Hermes Bloomfield-Gadêlha}
\affiliation{Department of Engineering Mathematics \& Bristol Robotics Laboratory, University of Bristol, BS8 1UB Bristol, UK}
\email[Hermes Bloomfield-Gadelha]{hermes.gadelha@bristol.ac.uk}

\begin{document}

%\input{example-content}

% PASA uses footnotes, not endnotes. \endnote in this template will behave like \footnote; and \printendnotes will not output anything.
% \printendnotes
%
%
%
%
%

\begin{abstract} % 250 words
Sperm modulate their flagellar symmetry to navigate through complex physico-chemical environments and achieve reproductive function. Yet it remains elusive how sperm swim forwards despite the inherent asymmetry of several components that constitutes the flagellar engine. Despite the critical importance of symmetry, or the lack of it, on sperm navigation and its physiological state, there is no methodology to date that can robustly detect the symmetry state of the beat in free-swimming sperm in 3D. How does symmetric progressive swimming emerge even for asymmetric beating, and how can beating (a)symmetry be inferred experimentally? Here, we numerically resolve the fluid mechanics of swimming around asymmetrically beating spermatozoa. This reveals that sperm spinning critically regularizes swimming into persistently symmetric paths in 3D, allowing sperm to swim forwards despite any imperfections on the beat. The sperm orientation in three-dimensions, and not the swimming path, can inform the symmetry state of the beat, eliminating the need of tracking the flagellum in 3D. We report  a surprising correspondence between the movement of sperm and spinning-top experiments, indicating that the flagellum drives ``spinning-top'' type rotations during sperm swimming, and that this parallel is not a mere analogy. These results may prove essential in future studies on the role of (a)symmetry in spinning and swimming microorganisms and micro-robots, as body orientation detection has been vastly overlooked in favour of swimming path detection. Altogether, sperm rotation may provide a foolproof mechanism for forward propulsion and navigation in nature that would otherwise not be possible for flagella with broken symmetry.

\end{abstract}

%\section{Significance} % 120 words
%\label{sec: sig}
%Sperm motility is critical for a successful reproduction. Spermatozoa is propelled by the active beating of a flagellum, yet it remains elusive how sperm can swim forwards despite the inherent asymmetry of the components that make up the flagellar engine. We tackle this problem by simulating the fluid mechanics around asymmetrically beating spermatozoa. We show that a novel spinning-top-like sperm rotation `filters out' any asymmetry to generate swimming paths that are always symmetric. This allows sperm to swim forwards despite any imperfection that may be present on the beat. Sperm drilling into the fluid provides a foolproof mechanism for forward propulsion in nature, as it would be otherwise impossible to `grow' a perfectly symmetric flagellar engine at the molecular level.

 \pdfoutput=1

\section{Introduction} 
\label{sec:intro}

\begin{figure*}[hbt!]
\centering
\begin{overpic}
[width=0.95\textwidth]{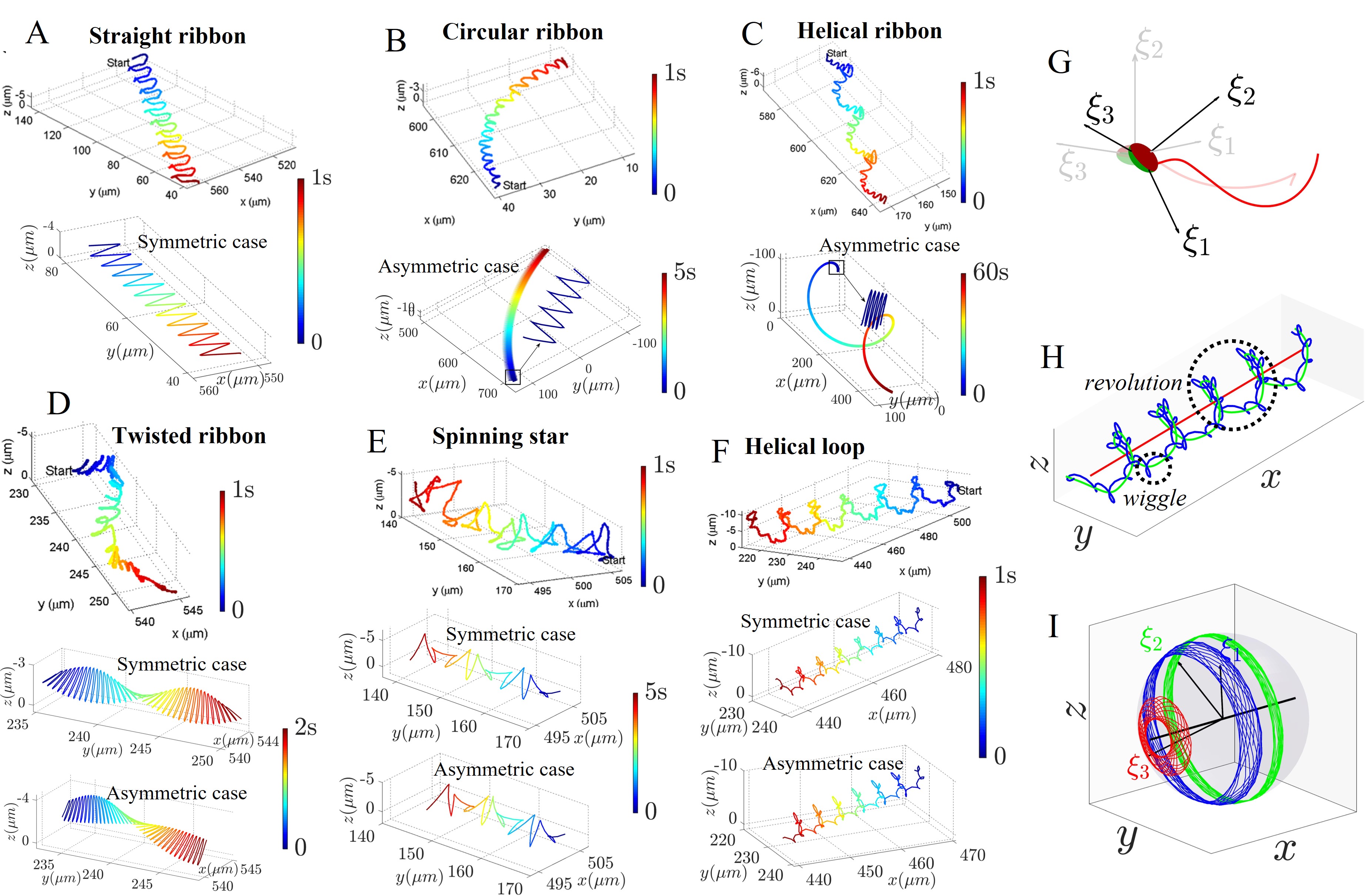}
\end{overpic}
\caption{\textbf{Comparison of experimental sperm trajectories and corresponding numerical reconstructions, together with diagrams of our numerical model.} Here illustrate 6 representative modes of sperm head track in the lab frame, with baselines on the top coming from the observations in \cite{daloglu2018label} and reconstructions in the middle and at the bottom generated by our numerical models. (\textbf{A})-(\textbf{F}) Trajectory modes of straight ribbon (SR), circular ribbon (CR), helical ribbon (HR), twisted ribbon (TR), spinning star (SS) and helical loop (HL), respectively.
Time sequence is represented by colors, and the simulated results are dimensionalized according to the bovine sperm arc length of 65$\mu m$ \cite{daloglu2018label, pesch2006structure, walker2020computer, magdanz2021impact, magdanz2020ironsperm} and the flagellum beating frequencies read from \cite{daloglu2018label}.
For some modes, such as SR, CR and HR, the representative trace patterns can only be generated by either symmetric or asymmetric virtual models, while for the others, the observed patterns can be reproduced by both of them.
(\textbf{G}) depicts snapshots of our numerical sperm at consecutive swimming moments in the lab frame, with the orthogonal vectors $\boldsymbol{\xi}_{1,2,3}$ indicating the time-varying orientations of the body frame. (\textbf{H}) tracks the simulated head center in the lab frame, where blue, green and red lines are separately for raw trajectory, average path and central axis, and the lab trace is characterized with local wiggle and global revolution. (\textbf{I}) condenses the orbits of $\boldsymbol{\xi}_{1,2,3}$ during lab swimming on a unit sphere and shows the head rotation of our virtual model. A black line is employed to indicate the head precession axis, around which the 3 orbits revolve.}
\label{image1: CompTraj ExpNum and NumModel}
\end{figure*}

Sperm navigate through female reproductive tract to fertilize the egg, encountering numerous hostile environments of viscous mucus, complex wall geometry, acid vaginal fluid and immune system, with small probability of success \cite{gaffney2011mammalian, tung2021co, gadadhar2021tubulin}. During this process, sperm flagellum plays a vital role providing motility and driving the cell forwards, via the emergence of both symmetric or asymmetric beating patterns \cite{gray1955movement, babcock2014episodic, jungnickel2018flagellar}. From centriole inner scaffold and its associated complex \cite{ounjai2012three, fishman2018novel, leung2021multi, chen2023situ, mali2023spokes} to dynein arrangement \cite{bui2012polarity, gibbons1961structural}, and ion channel distributions \cite{miller2018asymmetrically}, asymmetry is present throughout the structure of sperm flagellum. Nevertheless, asymmetric modulation of the beat is also critical for sperm capacitation, hyperactivation, signalling and chemotaxis, and essential during fertilization~\cite{mortimer1997critical, morgan2008tissue, gaffney2011mammalian, zaferani2021mammalian, jikeli2015sperm}. As such, waveform asymmetry is a critical proxy to distinguish different physiological states of the sperm flagellum. Despite the  importance of symmetry state detection of sperm flagella, there is no methodology to date that can robustly measure waveform asymmetry in free-swimming sperm in 3D (Fig.~\ref{image1: CompTraj ExpNum and NumModel}).

In the context of sperm swimming, it is generally accepted that symmetrical waveform leads to straight swimming trajectories, whilst asymmetrical beating patterns result in asymmetric swimming paths \cite{brokaw1974calcium, goldstein1977asymmetric, friedrich2010high, hansen2018spermq, gadelha2010nonlinear}. In an apparent contradiction, although asymmetry is intrinsic to the flagellar apparatus \cite{ounjai2012three, fishman2018novel, leung2021multi, chen2023situ, mali2023spokes, bui2012polarity, gibbons1961structural, miller2018asymmetrically,khanal2021dynamic}, three-dimensional (3D) sperm tracking experiments (Fig. \ref{image1: CompTraj ExpNum and NumModel}) show that the majority of sperm has a progressive swimming helical path with a globally straight forward direction \cite{su2012high, daloglu2018label, daloglu20183d}. Fig. \ref{image1: CompTraj ExpNum and NumModel} C-F (top row) depicts four representative categories of the experimental sperm head trajectories taken from \cite{daloglu2018label} that exhibit progressive swimming: helical ribbon (HR), twisted ribbon (TR), spinning star (SS) and helical loop (HL), representing $91.7\%$ of a total of 2133 tracked bovine sperm in 3D. Indeed, it has been long hypothesised in the literature that, despite the presence of waveform asymmetry, global forward motion is enabled by the out-of-plane beating component that drives sperm rotations as it swims \cite{rikmenspoel1960cinematographic, david1981kinematics, jikeli2015sperm, sartori2016curvature,zaferani2021rolling}--- though the exact mechanisms by which this could take place remain unexplored \cite{zaferani2021rolling}. Here, we test this hypothesis using mathematical modelling and simulation of free-swimming sperm driven by a symmetric and asymmetric beating flagellum in 3D. This reveals a novel spinning-top-like motion that regularizes sperm swimming into persistently symmetric swimming paths in 3D, with important consequences on the sperm swimming and empirical detection of flagellar waveform asymmetry, or symmetry, in sperm.

Numerical simulations in Fig.~\ref{image1: CompTraj ExpNum and NumModel} show that both symmetric and asymmetric waveforms recapitulate the diversity of experimental sperm trajectories in 3D \cite{daloglu2018label}, with progressive swimming modes, such as HR, TR, SS and HL, also in agreement with earlier studies \cite{jikeli2015sperm,smith2009boundary,sartori2016curvature}. This is in contrast with the simpler case of planar asymmetric waveforms which always lead to biased, curved trajectories~\cite{friedrich2010high} (Fig.~\ref{image1: CompTraj ExpNum and NumModel}, CR). The persistent progressive swimming shown in Fig.~\ref{image1: CompTraj ExpNum and NumModel} is also consistent with recent research by Zaferani et al.~\cite{zaferani2021rolling, zaferani2021mammalian} showing that progressive swimming is possible with asymmetric beating, which was attributed, once again, to the potential counteraction of the sperm rolling. Nevertheless, our numerical simulations in Fig.~\ref{image1: CompTraj ExpNum and NumModel} show that the symmetry state of the flagellum cannot be inferred from sperm trajectories alone, even if detected in 3D, as in \cite{daloglu2018label}, or if 3D flagellar beating information is available for free-swimming sperm at the lab frame of reference \cite{corkidi2008tracking, dardikman2020high, mojiri2021rapid, hansen2021multifocal}, as we further demonstrate in this study. In all, given the persistent symmetry of the swimming paths, how does the waveform asymmetry affect the sperm motion in 3D and how can this be detected experimentally? It is implausible that any source of asymmetry is simply `filtered out' from the system because of sperm rotations, according to the above hypothesis. In other words, waveform asymmetry must be manifested and detectable at some level during cell swimming, and this is indeed what we find here: the waveform asymmetry drives sperm in complex rotational orbits in 3D, thus far overlooked in the literature.

The complex interplay between body rotations and asymmetry is not exclusive to sperm swimming. Rolling disturbances may cause resonance and catastrophic yawing in missiles \cite{price1967sources, murphy2009spin, ananthkrishnan1999transient, cooper2012flight}, whilst rifle bullets obtain gyroscopic stability and improved accuracy with the appropriate spin \cite{chen2020aerodynamic, fresconi2012flight}. The physics behind the above rotating systems and sperm swimming is very distinct, yet similarities of motion patterns may indicate deeper mathematical connections among such disparate systems. This is indeed the case for dynamics of spinning tops described by Euler equations and the statics of elastic rods governed by Kirchhoff equations \cite{nizette1999towards, davies19933}. In the same spirit, we report here a surprising correspondence between simulations of sperm swimming and experimental tracks of spinning-tops \cite{spinning} (Fig. \ref{image4: Spinning top}), revealing that 3D sperm rotation is qualitatively similar to spinning-tops, and that this parallel is not a mere analogy. The sperm flagellum cycles around the head-tail junction and drives spinning-top-like rotations on the entire cell during sperm swimming, suggesting that a mathematical equivalence between the dynamical systems of these seemly unrelated motion types may exist.

\begin{figure*}[hbt!]
\centering
\begin{overpic}
[width=0.99\textwidth]{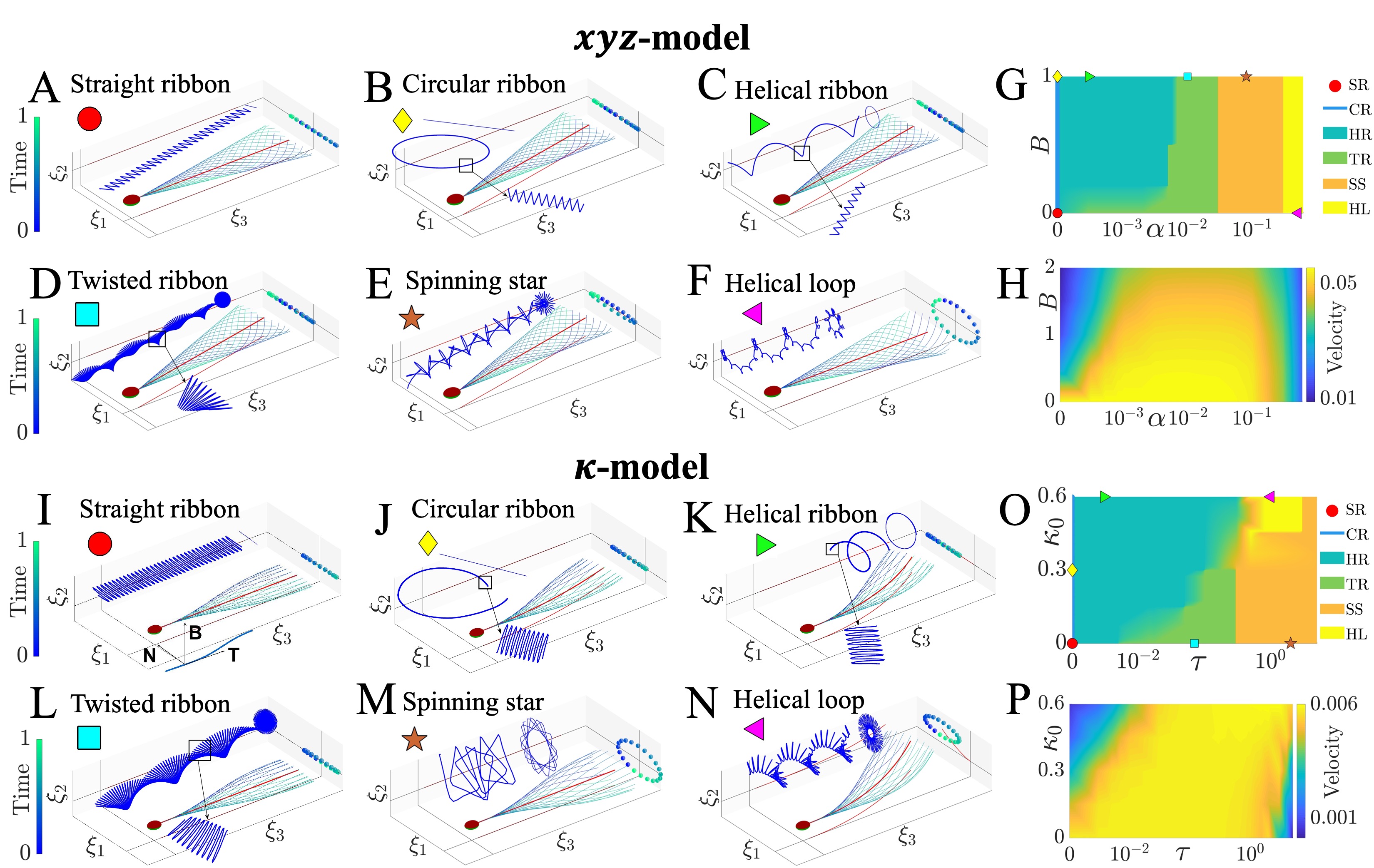}
\end{overpic}
\caption{\textbf{Virtual sperm model, lab trace classification and progressive velocity.} Flagellum waveform relative to the body frame for the $xyz$-model, (\textbf{A})-(\textbf{F}), and $\kappa$-model, (\textbf{I})-(\textbf{N}), are shown, together with the subsequent lab trajectories of sperm head center, the cases of which are denoted by different markers. 
The upside and downside of sperm head are distinguished by 2 different colors, and the flagellum positions over 1 beat cycle are color-encoded. Waveform asymmetry is indicated by the average flagellum shape in red, and the out-of-plane component can be identified from the projected point clouds.
6 representative trajectory modes varying with waveform asymmetry and rolling factors are displayed in (\textbf{G}) and (\textbf{O}), for the xyz-model using $k=2\pi$ and $\kappa$-model, respectively, where the exemplified cases given by (\textbf{A})-(\textbf{F}) and (\textbf{I})-(\textbf{N}) are marked. 
(\textbf{H}) and (\textbf{P}) Progressive velocity of the lab trajectory, for the $xyz$-model using $k=2\pi$ and $\kappa$-model, respectively. The linear speed is dimensionless and in units of flagellar arc length/ beat cycle.
}
\label{image2: NumModel TraceClass ProgVel}
\end{figure*}

\section{Results}
\label{sec:result}
We have conducted non-local microhydrodynamic simulations of free sperm swimming to elucidate the role of intrinsic waveform asymmetries on the resulting 3D sperm motion. We focus on the coupling between the out-of-plane component of three-dimensional helicoid waveforms \cite{ishijima1986flagellar, gadadhar2021tubulin, powar2022unraveling, jikeli2015sperm, kantsler2014rheotaxis}, which regulates sperm rotations in 3D through the waveform parameters $\alpha$ or $\tau$ (as detailed in Materials and Methods), and two types of static beating asymmetries observed experimentally: (i) a one-sided waveform shift~\cite{morgan2008tissue, khanal2021dynamic}, regulated by the $B$ parameter in the $xyz$-waveform model ($xyz$-model), in Fig. \ref{image2: NumModel TraceClass ProgVel} A-F, and (ii) a static curvature waveform bias~\cite{ bukatin2015bimodal, gong2021reconstruction,friedrich2010high,zaferani2021rolling}, captured by the average curvature $\kappa_0$ in the $\kappa$-waveform model ($\kappa$-model), in Fig.~\ref{image2: NumModel TraceClass ProgVel} I-N. A mathematical description of the models, estimations of parameter values, and the previous works that our numerical simulations are built on can be found in the Materials and Methods section.

\subsection{Symmetric and asymmetric beating patterns recapitulate experimental sperm swimming trajectories in 3D}
%\label{subsec: model recapitulate experiment}

Fig.~\ref{image1: CompTraj ExpNum and NumModel} and Table \ref{TabCompDynamic_HL} show excellent agreements with representative experimental sperm trajectories and dynamics~\cite{daloglu2018label, su2012high}, with both \textit{symmetric} and \textit{asymmetric} waveforms (Methods), highlighting the validity of the framework employed, and in further agreement with early studies \cite{gong2021reconstruction, su2013sperm, daloglu20183d, jikeli2015sperm, gadadhar2021tubulin, gallagher2019rapid, zaferani2021rolling, friedrich2010high}, for straight ribbon (SR), circular ribbon (CR), helical ribbon (HR), twisted ribbon (TR), spinning star (SS) and helical loop (HL). Simpler trajectory patterns, such as SR, CR and HR, can only be obtained by either symmetric (SR) or asymmetric (CR, HR) waveforms, whereas Fig.~\ref{image1: CompTraj ExpNum and NumModel} D-F show that TR, SS and HL modes are not exclusive to symmetric or asymmetric beating patterns. Asymmetric waveforms can generate almost indistinguishable swimming trajectories to the symmetric ones in 3D, thus both cases provide excellent agreement with experiments. That is, symmetric forward trajectories can be equally observed for both symmetric and asymmetric waveforms, even when the beat patterns are highly asymmetric, for both types of asymmetry (waveform shift or static curvature). Finally, in Table \ref{TabCompDynamic_HL} we compare our numerical results for the HL mode with experimental measurements of sperm motion from Ref.~\cite{su2012high}, showing once again excellent accordance, further validating our simulations, waveform models and parameter choices quantitatively.

\begin{table*}
\begin{threeparttable}
\caption{Comparison of numerical and experimental sperm dynamics for HL trajectory mode}\label{TabCompDynamic_HL}
\begin{tabular}{ c c c c c } \toprule
Data source  &Curvilinear velocity          &Helix radius      &Helix pitch              &Period of helix revolution            \\
     &($\mu$m/ s)           &($r$/ $\mu$m)       &($P$/ $\mu$m)      &(revolution/ s)     \\ \midrule
Numerical (xyz-model ($k=2\pi$))\tnote{a}  &43.0 -167.6   &2.1-2.4   &4.3-6.1   &1.7-8      \\ 
Numerical ($\kappa$-model ($k=2\pi$))   &43.6 -96.8   &2.2-2.5   &2.5-10.5   &0.3-2.7      \\ 
Experimental \cite{su2012high} & 68.8-129.4   &1.1-2.1  &2.7-4.7   &2.2-11.4        \\ \bottomrule
\end{tabular}
\begin{tablenotes}
\item[a] The raw numerical data is dimensionless such that sperm flagellum length is taken as 1 and time metric is beat cycle. To dimension the dynamic results, the length and beat frequency of human sperm appendage are taken as 55 $\mu$m and 10-20 Hz, respectively \cite{gaffney2011mammalian}.
\end{tablenotes}
\end{threeparttable}
\end{table*}

\subsection{Waveform rotation amplitude and asymmetry regulate the diversity of swimming paths in 3D}
%\label{subsec: anisotropy and asymmetry regulate the diversity}

Fig. \ref{image2: NumModel TraceClass ProgVel} A-F ($xyz$-model) and I-N ($\kappa$-model) show waveforms with different combinations of beating asymmetry ($B, \kappa_0$) and out-of-plane rotation amplitude ($\alpha, \tau$), together with the resultant diversity of swimming trajectories (insets): SR, CR, HR, TR, SS and HL. This highlights how small differences in waveform provoke large distinctions in swimming paths. Fine structures of the paths are recorded in Movies S1-S2, where intricate cusp formation and sharp turns are highlighted, as exemplified in Fig. \ref{image2: NumModel TraceClass ProgVel}N, and in particular, the trace patterns of HL mode exhibit local loops and global revolutions (Fig. \ref{image1: CompTraj ExpNum and NumModel}H) with opposed chirality, similar to tracks previously reported in \cite{mojiri2021rapid, daloglu2018label}. Note that the simulated swimming paths have been re-aligned to the $x$-axis above, for comparison purpose, as described in Fig. \ref{image1: CompTraj ExpNum and NumModel}H and Fig. \ref{image3: Param} A and B. Our simulations show that the overall direction of the swimming path is governed by both the waveform asymmetry and the out-of-plane component of the beat, as shown in Movie S3. As such, it may be possible that sperm can dynamically adjust these controls alone (waveform asymmetry and out-of-plane amplitude of rotation) to navigate in 3D in response to changes in the environment.

Fig.~\ref{image2: NumModel TraceClass ProgVel} G and O, and Fig.~S1, show the diversity map of trajectory-type in the asymmetry-rotation parameter space ($B-\alpha$ and $\kappa_0-\tau$). When the flagellar rotation amplitude is small ($\alpha,\,\tau$ low), the waveform asymmetry dictates the variations of swimming patterns. On the other hand, when the out-of-plane component is high ($\alpha,\,\tau$ high), the waveform asymmetry has negligible influence on the \textit{trajectory type}; with the exception of HL for the $\kappa$-model, which is only possible for large values of $\kappa_0$ and $\tau$ in Fig.~\ref{image2: NumModel TraceClass ProgVel}O. For example, TR switches to HR by increasing $B$, for $\alpha$ between $10^{-3}$ to $10^{-2}$, whilst for $\alpha$ larger than $10^{-2}$, the trajectory-type is independent of $B$, as similarly observed for the $\kappa$-model.

Fig. \ref{image2: NumModel TraceClass ProgVel} H and P, as well as Fig. S2, show the progressive velocity along the trajectory central axis (red line in Fig. \ref{image1: CompTraj ExpNum and NumModel}H) varying with asymmetry ($B$ or $\kappa_0$) and rotation amplitude ($\alpha$ or $\tau$) of the waveform. 
As expected, highest speeds appear for symmetric waveforms. However, even though flagellar asymmetry weakens the forward velocity, the out-of-plane component is able to increase any reduction in progressive speed caused by asymmetry, with increasing $\alpha$ or $\tau$, compensating in this way the detrimental effect that asymmetry has on progressive motion.
The influence of $\alpha$ and $\tau$ for both $\kappa$- and $xyz$-models are non-monotonic, indicating that an optimal level of waveform rotation amplitude exists for a given asymmetric waveform, to maximize the progressive speed.

\subsection{Waveform asymmetry is suppressed in the sperm swimming paths but manifested in the 3D orientation orbits}
%\label{subsec: spatial suppress}

\begin{figure*}[hbt!]
\centering
\begin{overpic}
[width=0.9\textwidth]{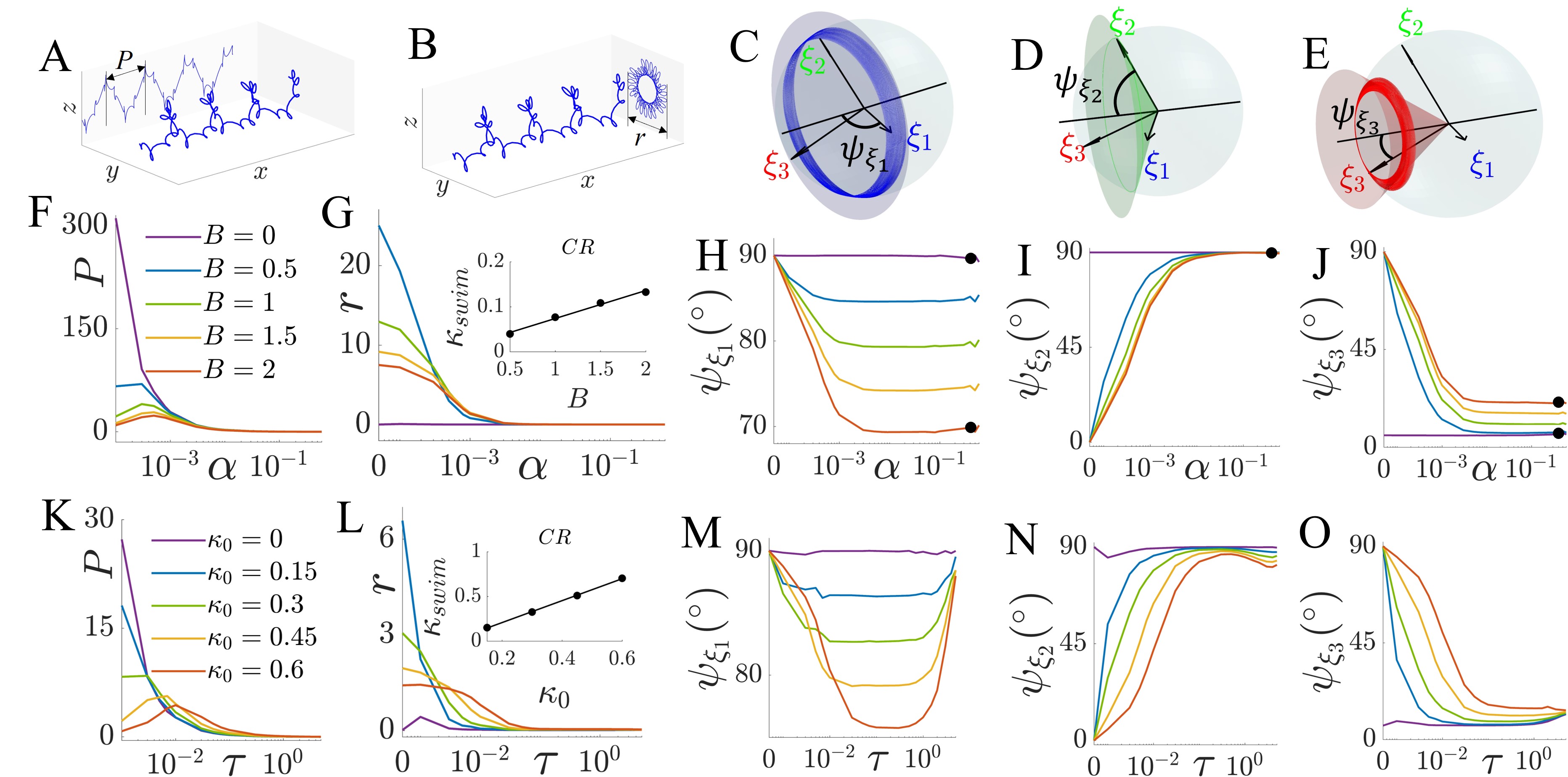}
\end{overpic}
\caption{\textbf{Parameterization of sperm locomotion, varying with waveform asymmetry and rolling.}
(\textbf{A}) and (\textbf{B}) Definitions of the longitudinal ($P$) and transverse ($r$) envelopes, based on the aligned sperm lab trajectory.
(\textbf{C})-(\textbf{E}) Definitions of the head tilt angles $\psi_{\boldsymbol{\xi}_{1,2,3}}$, the angles between the head rotational axis and the mean orbits of $\boldsymbol{\xi}_{1,2,3}$.
(\textbf{F})-(\textbf{J}) Parameterization of the xyz-model with $k=2\pi$, including $P$, $r$ and $\psi_{\boldsymbol{\xi}_{1,2,3}}$, where the inset demonstrates a linear relationship between trajectory curvature $\kappa_{swim}$ and waveform asymmetry in CR mode. The dots in (\textbf{H})-(\textbf{J}) mark the 2 cases shown in Fig. \ref{image6: flag beat in 3 frames}. 
(\textbf{K})-(\textbf{O}) Same results but for the $\kappa$-model.}
\label{image3: Param}
\end{figure*}

We measured the pitch and radius of the symmetric helical paths ($P$ and $r$), as well as the tilt angles of head orientation ($\psi_{\boldsymbol{\xi}_{1,2,3}}$) in the lab frame, as defined in Fig. \ref{image3: Param} A-E, and parameters displayed in Fig. \ref{image3: Param} F-O and Fig. S3. Both $P$ and $r$ are affected by the waveform asymmetry, and at the same time, regulated by the out-of-plane component of the beat. As $\alpha$ and $\tau$ increase and the waveform becomes more circular in the cross-section, the large pitch and radius in Fig. \ref{image3: Param} F, G, K and L drop, and ultimately decay to zero, regardless of the level of asymmetry, indicating a transition from large helical paths (small $\alpha, \tau$) to a linear forward movement (large $\alpha, \tau$), see Movie S4.
When $\alpha, \tau =0$, the characterised circular trajectory of the CR mode demonstrates a linear relationship between trajectory curvature $\kappa_{swim}$ and waveform asymmetry (insets of Fig. \ref{image3: Param} G and L, Fig. S3 B and G), consistent with observations by \cite{zaferani2021rolling, friedrich2010high}. Fig. \ref{image3: Param} F, G, K and L show that any distinction between the swimming paths due to the waveform asymmetry is lost as the out-of-plane component of the beat is increased (all curves collapse to zero), indicating that the flagellum rotation amplitude can indeed suppress the manifestation of the waveform asymmetry at the swimming path level whilst promoting global forward motion. %As expected for planar beating, both $xyz$ and $\kappa$-models follow a linear relationship between curvature of the swimming path $Kappa_{swim}$ and the wave asymmetry parameter ($B$,$\kappa_0$), as observed experimentally \cite{friedrich2010high}. 

The waveform asymmetry instigates complex rotational orbits in 3D (Fig. \ref{image3: Param} C-E, H-J and M-O). When $\alpha$ and $\tau$ are varied, the tilt angles of the sperm head orientation shown in Fig. \ref{image3: Param} H-J and M-O remain fairly constant for symmetric waveforms, whilst those for asymmetric waveforms are affected noticeably--- see comparisons provided in Movies S4 and S5, where the orbits of the head orientation vector $\boldsymbol{\xi}_2$ are regularized, in contrast to the wobbling traces of $\boldsymbol{\xi}_{1,3}$.
As $\alpha$ ($\tau$) increases, $\psi_{\boldsymbol{\xi}_{2}}$ and $\psi_{\boldsymbol{\xi}_{3}}$ tend to approach $90^{\circ}$ and $0^{\circ}$, respectively, but asymptote to slightly different angles depending on the level of waveform asymmetry. The head orientation of asymmetric cases, for the basis vectors $\boldsymbol{\xi}_{2,3}$, align more closely with the corresponding symmetric cases, as the flagellum rotation amplitude increases, indicating the suppression of rotation in two orientation directions as $\alpha$ and $\tau$ increases. Most importantly, the spherical orbits of $\boldsymbol{\xi}_1$ direction differ dramatically depending on the level of waveform asymmetry, even when the out-of-plane component of the beat ($\alpha$,$\tau$) is large (Fig. \ref{image3: Param} C, H and M). In the case of static asymmetric curvature in Fig. \ref{image3: Param} M, this distinction becomes less prominent for very large values of $\tau$, indicating an asymmetry-dependent effect on the dynamics of $\boldsymbol{\xi}_{1}$, as well as $\boldsymbol{\xi}_{3}$, when comparing Figs. \ref{image3: Param} J and O for large values of $\alpha$ and $\tau$.

\subsection{Sperm rotates like a spinning-top}
%\label{subsec: spinning-top}

The sperm head centre position (path trajectory) revolves around the central axis (Fig. \ref{image1: CompTraj ExpNum and NumModel}H and Movie~S4), whilst the sperm head orientation rotates around the central axis during its spherical orbit (Fig. \ref{image3: Param} E, J and O) with a precession motion, as depicted by the red nutating (wobbling) trajectory in Fig. \ref{image3: Param}E. In other words, the head long axis $\boldsymbol{\xi}_{3}$ revolves around the progressive swimming direction as the head spins around itself, defining the sperm head precession. To further quantify the sperm head precession, we define the angle between central axis (red line in Fig. \ref{image1: CompTraj ExpNum and NumModel}H) and head precession axis (black line in Fig. \ref{image1: CompTraj ExpNum and NumModel}I) as $\gamma$. Fig.~S5 shows the statistics of $\gamma$, where the medians and interquartile ranges of the angle vary closely around zero, with larger angles ranging between 10 and 30 degrees.  
The tendency of $\psi_{\boldsymbol{\xi}_3}$ declining towards $0^{\circ}$ (Fig. \ref{image3: Param} J and O, and Fig. S3 E and J) implies that a larger waveform rotation amplitude ($\alpha$ and $\tau$) leads to a head orientation in which the head long axis is almost parallel to the head precession axis and its progressive swimming direction. This is similar to the movement of a spinning-top. Movie~S4 presents typical precession and nutation movements of the sperm head long axis around the progressive direction (precession axis), while the head spins around its own longitudinal axis $\boldsymbol{\xi}_3$, in addition to oscillatory nutations, characterized by the wobbling movement of head long axes as it rotates/precesses around the progressive direction.  \\

\begin{figure*}[hbt!]
\centering
\begin{overpic}
[width=0.8\textwidth]{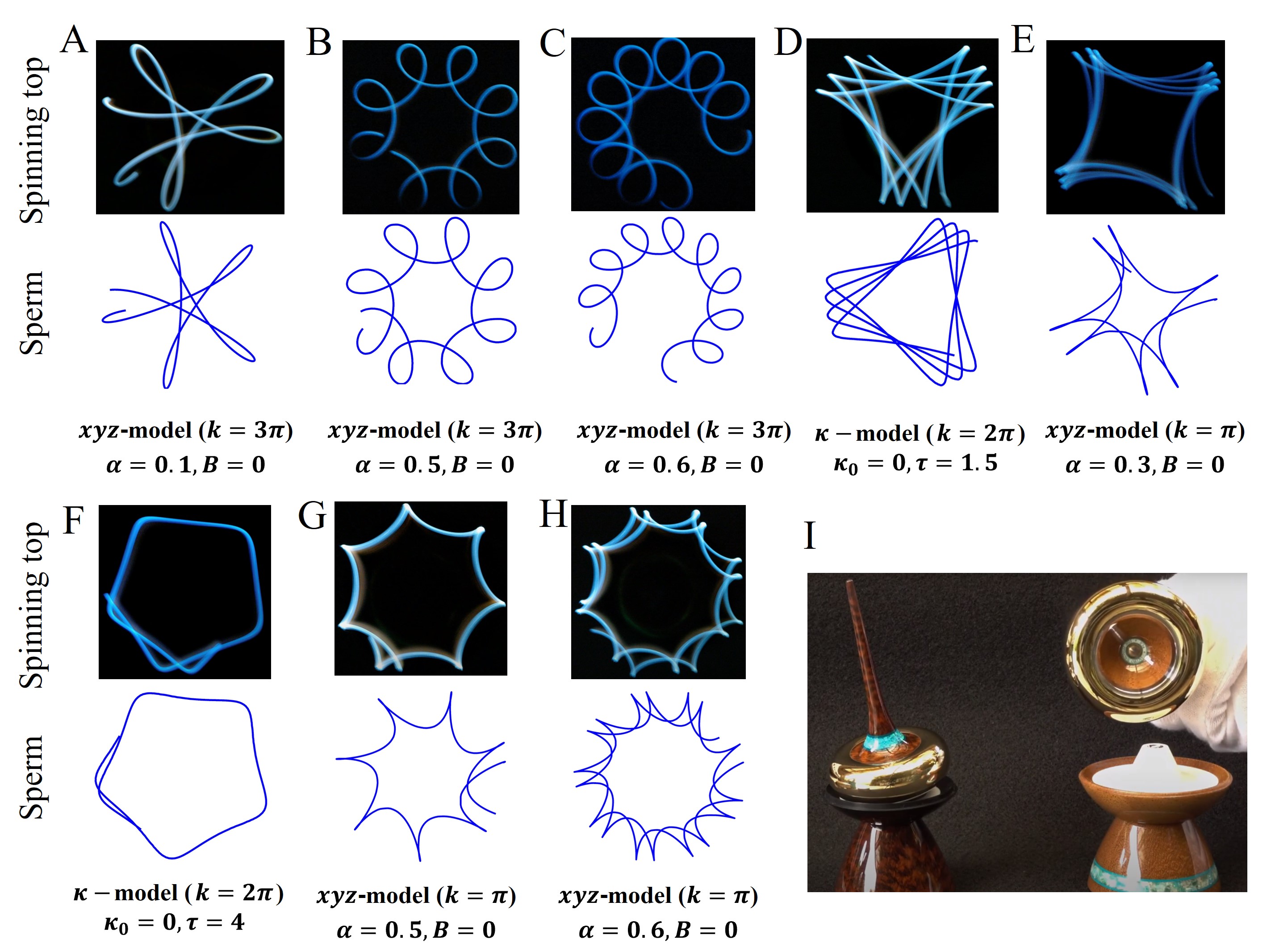}
\end{overpic}
\caption{\textbf{The orbits of spinning top are strikingly similar to the trajectory projections of sperm head center in the YZ plane (see Fig. \ref{image1: CompTraj ExpNum and NumModel}H), both characterized with local spirals or cusps and a global circle.} 
The upper graph in each column of (\textbf{A})-(\textbf{H}) is from experimental recordings of the spinning top self-rotation axis \cite{spinning}, and the lower graph is from our numerical simulations, with virtual model parameters listed at the bottom.
The chosen experimental orbits are generated by a specific top configuration \cite{TopConfig}, as shown in (\textbf{I}), where the center of mass is below the tip.}
\label{image4: Spinning top}
\end{figure*}

Fig. \ref{image4: Spinning top} shows the striking similarity between  observations of spinning-top orbits \cite{spinning}, obtained by tracking the spinning-top long axis, and the rotation patterns of the sperm head around the swimming directions ($yz$ projections of helical paths in Fig. \ref{image1: CompTraj ExpNum and NumModel}H). All trajectories are characterized by local loops (or cusps) following a global revolution around the centre (Fig. \ref{image4: Spinning top}). The remarkable similarity among such diverse trace patterns suggests that such qualitative comparison is not a mere analogy, and that an underlying equivalence may exist between the dynamical systems that govern these movements. Fig.~\ref{image4: Spinning top} shows the experimental orbits of spinning-tops with mixed precession and nutation movements. The spinning-top configuration in Fig. \ref{image4: Spinning top}I \cite{TopConfig} is referred as a `bottom-heavy' type of spinning-top, as its centre of mass is located below the spinning tip \cite{spinning, TopConfig}, resulting in the formation of outward-directed local loops. When the nutation movement is small, the local loops become less developed and ultimately degenerate into cusps, as shown in Fig. \ref{image4: Spinning top} D-H. In direct correspondence, sperm beatings with a lower wavelength, such as $k=\pi$, exhibit sharper turns in their projected trajectories leading to cusp formations, while large values generate loops. Waveform characteristics thus instigate spinning-top-like effects on the sperm swimming: sperm head nutation defines the type of helical path that emerges whilst regulated by the wavelength of the beat and the out-of-plane component. It is noteworthy that the sperm trajectories presented in Fig. \ref{image4: Spinning top} are for large flagellar rotation amplitudes ($\alpha$ and $\tau$), and traces with higher values of ($\alpha$, $\tau$) display a much denser appearance of loops and cusps.

\subsection{Waveform rotation inhibits asymmetry in the sperm orientation orbital cycle}
%\label{subsec: temporal inhibit}

\begin{figure}[hbt!]
\centering
\begin{overpic}
[width=1\textwidth]{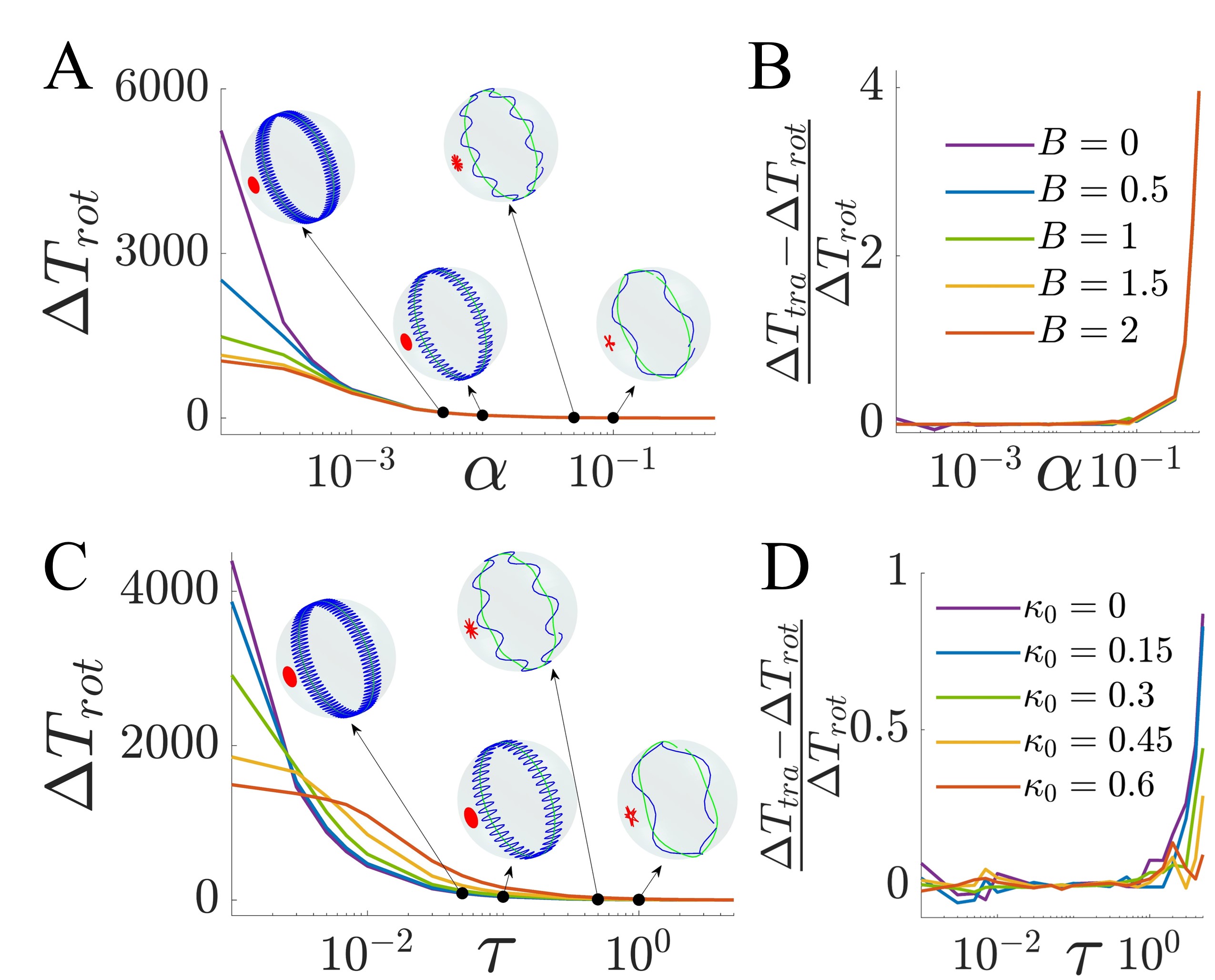}
\end{overpic}
\caption{\textbf{Cyclic periods of spherical head rotation ($\Delta T_{rot}$) and lab trace revolution ($\Delta T_{tra}$), in function of waveform asymmetry and rolling.}
(\textbf{A}) Spherical periods of the xyz-model using $k=2\pi$.
Insets exemplify the spherical orbits within 1 $\Delta T_{rot}$, the spatial frequency of which decreases when $\alpha$ increases. (\textbf{B}) Relative deviation between the trace revolution cycle and the spherical rotation period, for the xyz-model using $k=2\pi$. 
(\textbf{C})-(\textbf{D}) Same results but for the $\kappa$-model.
}
\label{image5: Rot Period}
\end{figure}

The time sperm takes to complete one period along its helical path (Fig. \ref{image1: CompTraj ExpNum and NumModel}H) is defined as $\Delta T_{tra}$, and the time head orientation vectors complete one spherical orbit (Fig. \ref{image1: CompTraj ExpNum and NumModel}I) is defined as $\Delta T_{rot}$, in units of beat cycle. Fig. \ref{image5: Rot Period} A and C, and Fig.~S4 A and C, show the orbital period of the head rotation, $\Delta T_{rot}$, as a function of $\alpha,\tau$ for distinct waveform asymmetries ($B,\kappa_0$). 
The smaller $\Delta T_{rot}$ period is, the faster the angular speed of head rotation will be. When $\alpha,\tau$ are small, $\Delta T_{rot}$ can be as large as thousands of beat cycles, depending on the level of asymmetry. 
However, as $\alpha,\tau$ increase, $\Delta T_{rot}$ decreases, regardless of the magnitude of the waveform asymmetry, with all cases collapsing into a fast sperm rotating mode, dominated by the out-of-plane component of the waveform.
For large $\alpha,\tau$, the orbital revolution is much faster (details see Movies S6, S7 and S8), yet characterised by a smoother wobbling movement and a lower spatial frequency, as depicted by the insets showing the spherical orbit for one revolution. As such, large waveform rotation amplitude inhibits the effect of waveform asymmetry on the sperm orientation orbital cycle.

Fig. \ref{image5: Rot Period} B, D, and Fig. S4 B, D, show the relative deviations between head trajectory helical period $\Delta T_{tra}$ and head rotation orbital period $\Delta T_{rot}$. For most of the range of $\alpha,\tau$ studied, the deviation is either zero or very small, indicating that head trajectory and rotation revolution follow similar behaviours. 
For instance, the decreasing $\Delta T_{rot}$ caused by the increasing $B$ ($\kappa_0$) at small $\alpha$ ($\tau$) implies a shorter period of head rotation movement, as well as a faster revolution of the translation trajectory, for a quasi-planar waveform with more asymmetry.
An apparent discrepancy between the periods of translation and rotation occurs when the waveform out-of-plane component is very large. Movie S7 shows an example of sperm motion where the revolutions of the head translation and spherical orbit are in synchrony, with zero deviation and $\alpha=0.05$, in contrast with Movie S8, in which head spinning is faster than the head trajectory revolution, with a deviation of 3.2 beat cycles and $\alpha=0.5$.

\subsection{Quantification of waveform asymmetry in free-swimming sperm requires detection of sperm head orientation in 3D}
%\label{subsec:asy detect}

\begin{figure*}[hbt!]
\centering
\begin{overpic}
[width=0.9\textwidth]{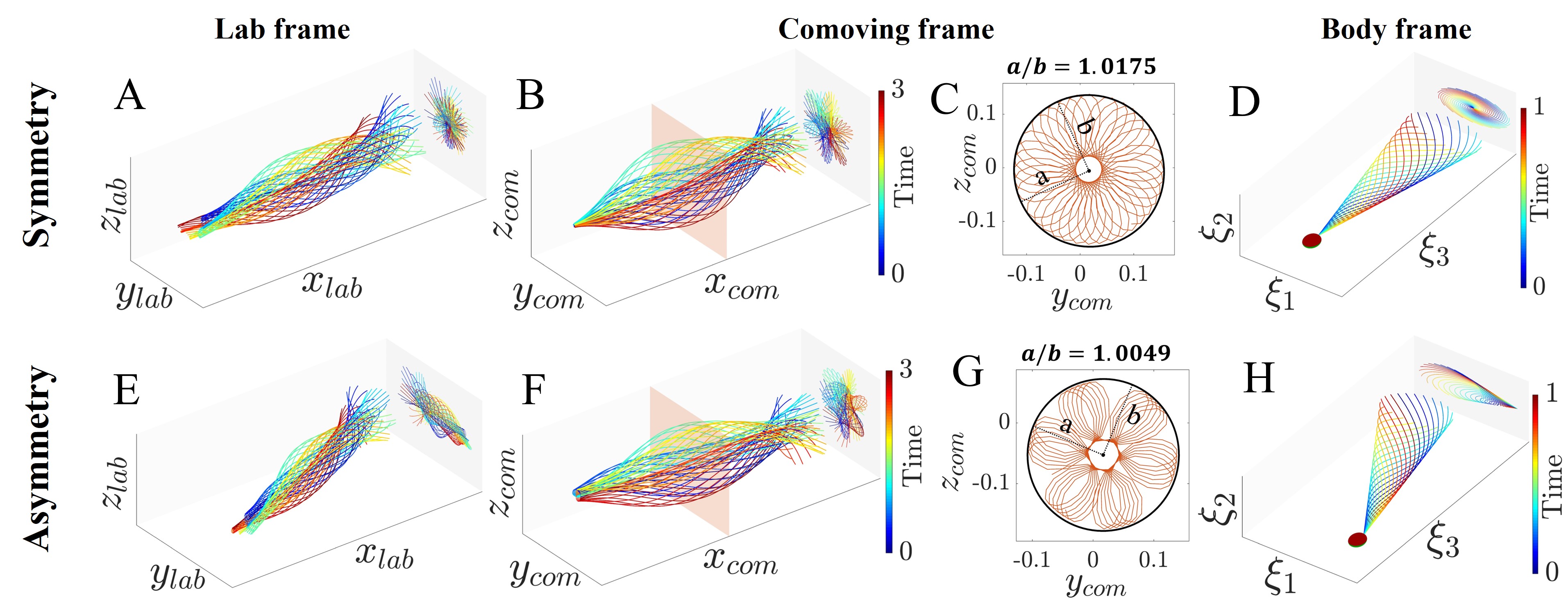}
\end{overpic}
\caption{ \textbf{Flagella beating in the lab, comoving and body frames of reference.}
The symmetry and asymmetry cases are marked in Fig. \ref{image3: Param} (\textbf{H})-(\textbf{J}) with black dots. 
(\textbf{A}) and (\textbf{B}) Flagellum beats in the lab and comoving frames, respectively, within 1 trace revolution period. The red plane indicates the approximate position of the mid-flagellum point, whose projected trace over revolutions is shown in (\textbf{C}). The black 'ellipse' obtained via PCA captures the three-dimensionality of the mid-flagellar track, and the ratio of the major axis to the minor axis, $a/b$, equals to 1.0175.
(\textbf{D}) Corresponding waveform in the body frame is symmetric.
Time sequence is represented by colors, in units of beat cycle.
(\textbf{E})-(\textbf{H}) Same results but for an asymmetric case, whose comoving 'ellipse' shows a ratio equal to 1.0049.
}
\label{image6: flag beat in 3 frames}
\end{figure*}

\begin{figure}[hbt!]
\centering
\begin{overpic}
[width=0.99\textwidth]{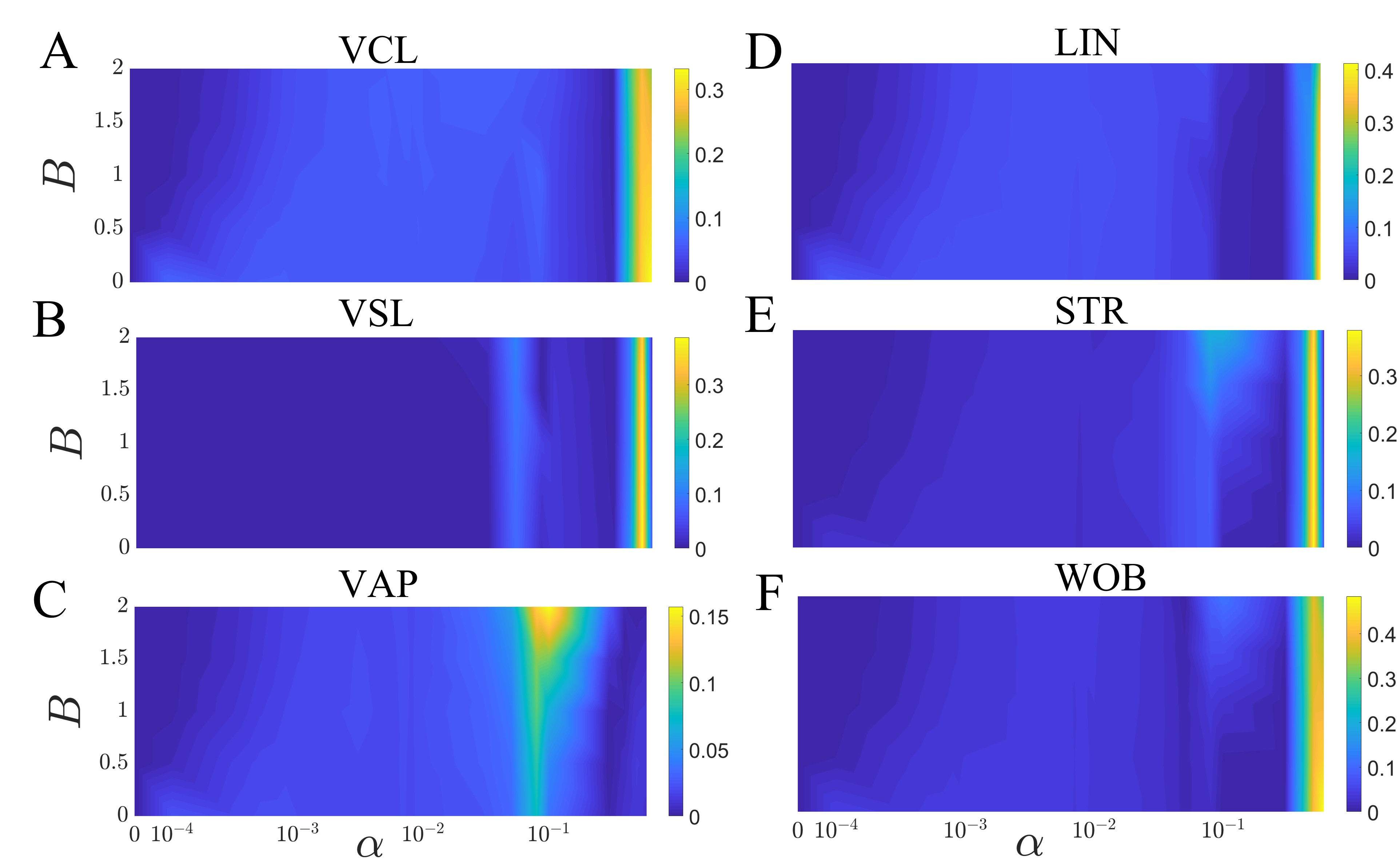}
\end{overpic}
\caption{ \textbf{Kymographs of the relative deviations between 2D and 3D CASA parameters, based on the xyz-model using $k=2\pi$.} 
(\textbf{A}) curvilinear velocity (VCL), (\textbf{B}) straight-line velocity (VSL), (\textbf{C}) average-path velocity (VAP), (\textbf{D}) linearity (LIN, equal to VSL/VCL), (\textbf{E}) straightness (STR, equal to VSL/VAP) and (\textbf{F}) wobble (WOB, equal to VAP/VCL) \cite{alquezar2019opencasa}, varying with $\alpha$ and $B$.}
\label{image7: CASA dev}
\end{figure}

The waveform asymmetry is not detectable from translational motion of the sperm head and flagellum in 3D at the lab frame of reference. As shown in the above sections, sperm rotation `filters-out' waveform asymmetry from linear translations and subsequent swimming paths. This is further illustrated in Fig. \ref{image6: flag beat in 3 frames}, which shows flagellar beating relative to the laboratory frame of reference (A, E), comoving frame of reference (B, F), and body frame of reference (D, H), for both symmetric and asymmetric waveforms. The beating patterns in the comoving frame (Fig.~\ref{image6: flag beat in 3 frames} B and F) are obtained from those in the lab frame such that one observes the flagellum translating with sperm head along the progressive direction (but not rotating with the sperm head), with projections on the $yz$-plane exhibiting symmetric distributions of the waveform tracers around the central swimming axis for both symmetric and asymmetric cases (Fig.~\ref{image6: flag beat in 3 frames} B and F). The principal components of the mid-flagellar tracers in Fig. \ref{image6: flag beat in 3 frames} C and G show that the fitted envelopes by ellipses have semi-axes ratio ($a/b$) close to one, see SI for details on principal component analysis (PCA). All flagellar motions in Fig. \ref{image6: flag beat in 3 frames} C and G appear to be symmetric. The symmetry state of the waveform is indistinguishable at both lab and comoving frames of reference. Indeed, we show here that asymmetry, or indeed symmetry, of the waveform is only detectable with full knowledge of the 3D head orientation, as displayed in Fig. \ref{image3: Param} H-J, where black markers denote the simulations depicted in Fig.~\ref{image6: flag beat in 3 frames} showing sizeable distinctions in angular differences between the asymmetric and symmetric cases. Detection of both head orientation and translations in 3D is thus ideal, as it allows reconstruction of the true beating of the flagellum, relative to the body frame of reference, from which the waveform symmetry state can be detected. The body frames of reference in Fig.~\ref{image6: flag beat in 3 frames} D and H clearly distinguish asymmetric from symmetric beatings.

As shown in Fig.~\ref{image1: CompTraj ExpNum and NumModel} D-F, it is challenging to distinguish asymmetric and symmetric beating patterns by inspecting sperm head trajectories alone, or even waveform trajectories in 3D (Fig.~\ref{image6: flag beat in 3 frames}). Fig. \ref{image3: Param} H, M, and Fig. S3 C, H, however, reveal that 3D angular movements of sperm head alone can distinguish asymmetric and symmetric waveforms without requiring waveform tracers nor head trajectories in 3D, even for large values of $\alpha,\tau$. When the waveform is symmetric ($B=0,\kappa_0=0$), the head orientation relative to the head precession axis, quantified by $\psi_{\boldsymbol{\xi}_{1,2,3}}$, does not change with $\alpha$ or $\tau$, and deviations from these angles are associated with the magnitude of the waveform asymmetry. As such, the 3D orientations in Fig.~\ref{image3: Param} can be used alone as a lookup table to infer waveform asymmetry. For the symmetric cases, $\psi_{\boldsymbol{\xi}_{1}}=\psi_{\boldsymbol{\xi}_{2}}=90^{\circ}$, and $\psi_{\boldsymbol{\xi}_{3}} = 7^{\circ}$, $5.5^{\circ}$, $3.6^{\circ}$ and $6.6^{\circ}$, respectively, for the $xyz$-model with $k=\pi, \ 2\pi,\ 3\pi$, and the $\kappa$-model. Movie S4 further shows how the orientation orbits of asymmetric waveforms vary with $\alpha$, in comparison with the symmetric cases in  Movie~S5.

\subsection{CASA parameters cannot detect waveform asymmetry of 3D beating spermatozoa}

Computer-assisted sperm analysis (CASA) systems are used in clinical settings to assess sperm motility from 2D tracers~\cite{mortimer2015future, world2010laboratory}. Here we present evaluations of the so-called CASA parameters of our numerical 3D sperm trajectories, namely curvilinear velocity (VCL), straight-line velocity (VSL), average-path velocity (VAP), linearity (LIN, equal to VSL/VCL), wobble (WOB, equal to VAP/VCL) and straightness (STR, equal to VSL/VAP) \cite{alquezar2019opencasa}. Figs. S6 and Fig.~\ref{image7: CASA dev} display the generalised 3D CASA parameters and the deviation between 2D projections of the 3D movements and the 3D results, respectively, as a function of waveform asymmetry and out-of-plane component of the beat.  The effect of waveform asymmetry is nearly indistinguishable in Figs. S6 and Fig.~\ref{image7: CASA dev}. CASA parameters are only weakly affected by the waveform asymmetry $B$, and thus they are unable to detect asymmetry, as also expected from the persistently symmetric swimming paths presented previously in Fig. \ref{image1: CompTraj ExpNum and NumModel}-\ref{image3: Param}, \ref{image6: flag beat in 3 frames}.

Although the 2D CASA parameters deviate only weakly from their 3D counterparts for most of the parameter space in Fig.~\ref{image7: CASA dev}, the relative deviation can be as high as $33.21\%$, $38.68\%$, $15.69\%$, $41.26\%$, $37.94\%$ and $48.40\%$, respectively, for VCL, VSL, VAP, LIN, WOB and STR. The absolute difference between 2D and 3D for VCL can reach up to $28.71\mu$m/s, a value that surpasses, for example, the threshold to distinguish slow from rapid progressive motility ($25 \mu$m/s), according to the WHO guidance \cite{world1999laboratory}, and constitutes approximately $1/5$ of the speed used for hyperactivated motility ($150\mu$m/s)~\cite{mortimer1998effect}. As such, 2D CASA measurements of 3D sperm trajectories may introduce unknown inaccuracies on sperm motility assessment using CASA parameters.

\section{Discussion}
\label{sec:discuss}

We have conducted numerical simulations of sperm swimming hydrodynamics to elucidate the role of intrinsic waveform asymmetry on the resulting 3D swimming movements. Numerical simulations were verified against experimental swimming observations and measurements~\cite{daloglu2018label, su2013sperm, gong2021reconstruction,su2012high,friedrich2010high, zaferani2021rolling}, with waveform model parameters estimated directly from experiments \cite{ishijima1986flagellar, miki2013rheotaxis, morgan2008tissue, khanal2021dynamic, bukatin2015bimodal, gong2021reconstruction, mojiri2021rapid, hansen2021multifocal}. Our study revealed a complex interplay between flagellar beat asymmetry and the sperm motions in 3D. Counterintuitively, we showed that the waveform asymmetry is not manifested in swimming path patterns, rather it impacts the three-dimensional head rotations in a complex manner. The swimming trajectories of both symmetrical and asymmetrical flagellar beatings are persistently symmetric. As such, 3D sperm trajectories alone cannot inform the symmetry state of beating patterns. Most interestingly, the waveform asymmetry information is `stored' in the head orientation dynamics. Indeed, the flagellar beat is a 3D helicoid that continuously cycles, rotating around a centre point, so that sperm head rotation (driven by this flagellar motion) depends directly on the level of asymmetry of the beating helicoid. Waveform asymmetry deviates the head orientation during motion (Movies S4-S5), particularly altering the relative angles between the head basis vectors and the head rotational axis (Fig. \ref{image3: Param} C-E, H-J and M-O, Fig. S3 C-E and H-J). We have showed that 3D head orientation alone is sufficient to scrutinize whether a given flagellar beat is symmetric or not. This may prove critical in future empirical studies, as 3D body orientation detection of microorganisms has been vastly overlooked in favour of trajectory detection in the literature.

The sperm flagellum apparatus possess an umbrella of intrinsic asymmetric components spanning from the molecular to micron level, including molecular motors, radial spokes and elastic linkers, microtubules, outer-dense-fibres, centrioles, basal components and ion channels \cite{ounjai2012three, fishman2018novel, leung2021multi, chen2023situ, mali2023spokes, bui2012polarity, gibbons1961structural, miller2018asymmetrically,khanal2021dynamic}, to name a few. But how does sperm achieve forward swimming with such intrinsically asymmetric beating flagella? We have shown that forward swimming motion is not hindered by waveform asymmetry, due to the regularising role of the rotational motion arising from the moment balance. This allows generic flagellar apparatus to propel cells forwards regardless of any `imperfection' that may drive the flagellar beat in an asymmetric manner. The rotational motion of the sperm flagellum thus provides a foolproof mechanism for forward propulsion in nature, which could be potentially critical during the evolution of these cell appendages while achieving biological function - it would be an impossible task to grow a `perfect' flagellar apparatus with exactly symmetric molecular components at every level. Our results suggest that imperfections of the flagellar beat would not dramatically influence their ability to swim forwards for 3D beating patterns.

The sperm's capacity to persistently swim in a straight helical path despite the waveform asymmetry, however, does not hinder their ability to steer and navigate in 3D. Different levels of waveform asymmetry, modulated by flagellar rotation, leads to helical paths in different directions in space (Movie S3). This may allow sperm to use asymmetric waveform controls to navigate in 3D. This can be achieved by simply tuning the waveform asymmetry and the out-of-plane component, without risking to `trap' itself into circular swimming paths, as observed for asymmetric planar waveforms \cite{friedrich2010high, zaferani2021rolling, miki2013rheotaxis, woolley2003motility}. 
Asymmetric modulation of the beat is also an important proxy used to inform sperm capacitation and hyperactivation \cite{mortimer1997critical, morgan2008tissue, gaffney2011mammalian, zaferani2021mammalian}, physiological states that are critical for fertilization. Our results indicate that it would be a challenging task to identify sperm hyperactivity, using the sperm's path asymmetry as a proxy, even if recorded in 3D, given the persistent symmetric characteristics of sperm trajectories for asymmetric beating we observe.
In all, the waveform asymmetry amplifies the diversity of helical swimming paths (Fig. \ref{image2: NumModel TraceClass ProgVel} G and O, Fig. S1), and due to the sperm rotations, detrimental effects on the swimming speed can be circumvented (Fig. \ref{image2: NumModel TraceClass ProgVel}H, Fig. S2).

Comparison between simulations and experimental sperm trajectories showed that both symmetric and asymmetric waveform models were able to reproduce the observed trajectory patterns (Fig. \ref{image1: CompTraj ExpNum and NumModel} D-F). As a result, symmetry of the waveform cannot be uniquely inferred from observations of swimming trajectories alone, and likewise, waveform model comparison with experiments, at sperm trajectory level, provides insufficient information to scrutinise model closeness to experiments. This indicates that head centre trajectories should be considered together with 3D head orientation dynamics for quantitative comparisons between free-swimming sperm experiments and model predictions~\cite{rossi2017kinematics}. This is particularly relevant, as comparisons between experiments and theory have been largely limited to the head trajectory level, thus calling for a reevaluation of the generally expected symmetry of both flagellar beat and swimming paths in 3D.

We have demonstrated that even 3D detection of the flagellum waveform taken at the laboratory frame of reference may provide insufficient information to scrutinise the presence of asymmetry in the beat (Fig. \ref{image6: flag beat in 3 frames}). In this case, detection of head translations, in conjunction with  head rotations, may be required for empirical inference of beat asymmetries, whilst also allowing the reconstruction of the ``true'' flagellar beat relative to the body frame of reference---the flagellar movement as viewed from a fixed point of reference located at the sperm head, which translates and rotates with the sperm (Fig. \ref{image1: CompTraj ExpNum and NumModel}G).
This is due to the fact that although the shape of the flagellum is the same in both laboratory and body frames of reference, they differ dramatically in location and orientation from each other. Mathematically, the unknown translations and rotations of the body frame for a prescribed waveform can be obtained by solving a well-posed momentum balance system of equations (section \ref{sec:meth}), known as the mobility problem in the low Reynolds number hydrodynamics~\cite{gallagher2018meshfree}. The inverse problem, on the other hand, of finding body frame movement from the lab frame flagellar beat, is not well-posed, as the flagellum centreline does not carry orientation information of its local basis vectors in relation to the body frame~\cite{rossi2017kinematics}. Hence, 3D flagellar tracking without direct body orientation detection may not fully inform how the flagellum beats at the body frame. Indirect inference of the body orientation from observed flagellar path in 3D has been employed as an alternative \cite{dardikman2020high, ishijima1992rotational}, though more research is needed to determine  
whether this approach can confidently resolve the complex rotational movement of the sperm body frame in 3D. We hope that these results will motivate further advances on high-precision and direct measurements of microorganisms' body orientation in 3D \cite{rossi2017kinematics,corkidi2022human}.

The sperm head rotational dynamics share very similar characteristics with the kinematics of spinning-tops (Fig. \ref{image3: Param}E), with the emergence of both precession and nutation movements. It is not the first time that a parallel between seemly unrelated systems was found with spinning-top dynamics. The celebrated Kirchhoff equations for the statics of elastic rods share an intimate relation with the governing dynamics of spinning rigid bodies~\cite{nizette1999towards, davies19933}, despite the very different physics involved. This may also be the case for sperm swimming in 3D. Fig.~\ref{image4: Spinning top} compares the trajectory patterns of sperm head center and observed orbits for bottom-heavy spinning-tops. The similarity between the trace patterns is striking, and reveals that indeed 3D sperm swimming is qualitatively similar to spinning-tops, and that this parallel is not a mere analogy. This remarkable similarity is despite the very distinct physics that govern the hydrodynamics of rotating sperm at the micro-scale and the inertial dynamics of spinning tops at the macro-scale \cite{goldstein2013classical, wittenburg2013dynamics}.
Fig. \ref{image4: Spinning top} suggests that a mapping between these systems may exist, in which waveform characteristics could instigate spinning-top-like effects on the resulting sperm dynamics, and that a potential mathematical equivalence between these seemly unrelated motion types is possible, including subsequent connection with static configurations of Kirchhoff rods~\cite{nizette1999towards, davies19933}.

In the context of experimental and clinical studies focusing on the assessment of sperm motility and hyperactivation, we observe that 2D measurements of 3D movements may oversimplify the true sperm swimming motion. Specifically, Fig. \ref{image7: CASA dev} and Fig. S6 show that the so-called CASA parameters cannot distinguish symmetric from asymmetric waveforms for spermatozoa swimming in 3D. Large errors may arise from tracking 2D projections of 3D sperm movements. This is particularly important as current CASA measurements are restricted to 2D visualizations of the sperm swimming trajectories due to the limitations of imaging techniques \cite{mortimer1997critical, mortimer2000casa, corkidi2022human, zaferani2021rolling}, and thus carry subsequent challenges on the potential misclassification of sperm motility \cite{guerrero2011strategies}.

Our numerical study on the role of waveform asymmetry on the resulting 3D sperm swimming has several limitations: we focused on only two generic types of empirically observed waveform asymmetries ~\cite{morgan2008tissue, khanal2021dynamic, bukatin2015bimodal, gong2021reconstruction}, one-sided waveform shifts and static waveform curvatures, but other types of waveform asymmetry may exist, in both static and dynamic forms \cite{khanal2021dynamic, gadadhar2021tubulin}, such as second-harmonics used to steer sperm in 2D \cite{saggiorato2017human}, and planar beating inclined to the plane of flattening of the sperm head \cite{woolley2003motility, smith2009human}. Our work only considers presence of the fundamental beating mode---other harmonics are equally observed in experiments, although they are manifested with much lower amplitudes \cite{meng2021conditions, saggiorato2017human, smith2009bend}. We only accounted for beating patterns observed in low viscosity fluids, and neglected any sperm interaction with nearby walls and boundaries, which are well known to instigate boundary accumulation of sperm \cite{ishimoto2015fluid, smith2009human, woolley2003motility}. The mathematical framework and analysis developed here, however, can be easily generalised in future studies focusing on these elements. We also considered prescribed waveform models and, as such, the flagellar shape does not emerge spontaneously from the collective behaviour of molecular motors \cite{cass2023reaction, oriola2017nonlinear}. Despite the number of simplifications invoked, our results shed new light on the fundamental importance of waveform asymmetry, the complex 3D rotational motion of spermatozoa, and the diversity of persistently symmetric swimming patterns, despite any intrinsic asymmetry that may exist on the beat. We hope this work will instigate future research on the role of asymmetry on cell motility and rotational motion of microorganisms, waveform tracking in 3D, sperm motility, and artificial swimmers.

\section{Materials and Methods} 
\label{sec:meth}

\subsection{Numerical simulations of sperm swimming in 3D}

We exploit a meshfree approach using the Regularized Stokeslet method (RSM) by Cortez-Fauci-Medovikov\cite{cortez2005method} to solve the non-local low Reynolds number hydrodynamics of sperm swimming. The RSM has been extensively studied and validated in the literature \cite{cortez2005method, rodenborn2013propulsion, jung2007rotational, ainley2008method, smith2009boundary, walker2019filament}, and we use the novel nearest-neighbor discretization method developed by Gallagher-Smith \cite{gallagher2018meshfree, gallagher2021art, smith2018nearest} for efficient computations of the non-local flow fields. Gallagher-Smith method offers model simplicity and versatility, and has been optimized and validated for free-swimming problems, more details can be found in \cite{smith2018nearest}, including a didactic Matlab implementation of the method. By invoking total momentum balance, this framework provides the free-swimming motion of a spermatozoon, relative to the laboratory fixed frame of reference (lab frame), by prescribing the beating pattern of the flagellum relative to the body fixed frame of reference (body frame), i.e. the reference frame that translates and rotates with the sperm head (Fig. \ref{image2: NumModel TraceClass ProgVel}). The microscale flow velocity at a spatial point $\boldsymbol {x}$, driven by a regularized force $\phi ^{\epsilon} (\boldsymbol{x}-\boldsymbol{X}) \cdot \boldsymbol{f}$ at the location $\boldsymbol{X}$, can be represented as $\boldsymbol{u}= \boldsymbol{G} ^{\epsilon} (\boldsymbol{x}, \boldsymbol{X}) \cdot \boldsymbol{f}$, where 
$\phi ^{\epsilon} (\boldsymbol{x}-\boldsymbol{X})=(15\epsilon ^4)/[8\pi(r^2+\epsilon^2)^{7/2}]$ is the cutoff function, and
$\boldsymbol{G} ^{\epsilon}=[(r^2+2\epsilon^2)\boldsymbol{I}+\boldsymbol{r}\boldsymbol{r}]/(8 \pi r_{\epsilon}^3) $ is the regularized Stokeslet, with $\boldsymbol{r}=\boldsymbol{x}-\boldsymbol{X}$, $r=|\boldsymbol{r}|$, $\epsilon$ is the regularization parameter, and $r_{\epsilon}=\sqrt{r^2+\epsilon^2}$ \cite{cortez2005method, jung2007rotational}. We describe the laboratory frame coordinates of the sperm as $\boldsymbol{x}=\boldsymbol{x}_0+\boldsymbol{R} \cdot \boldsymbol{\xi}$, where $\boldsymbol{x}_0$ is the origin of the body frame, i.e. head centre, $\boldsymbol{R}=[\boldsymbol{\xi_1}, \boldsymbol{\xi_2}, \boldsymbol{\xi_3}]$ is director basis capturing the orientation of the body frame, and $\boldsymbol{\xi}$ the body frame coordinates of the flagellum shape (Fig. \ref{image2: NumModel TraceClass ProgVel}). The sperm velocity in the lab frame can be expressed as the boundary integral over the body surface $\partial D$,
\begin{equation}
\begin{aligned}
   \label{equ:1} 
   \boldsymbol{U}+\boldsymbol{\Omega}\times(\boldsymbol{x}-\boldsymbol{x}_0)+\boldsymbol{R}\cdot \Dot {\boldsymbol \xi} = \iint_{\boldsymbol X \in \partial D} \boldsymbol{G} ^\epsilon (\boldsymbol x, \boldsymbol X) \cdot \boldsymbol{f} (\boldsymbol X) d\boldsymbol X, %\cite{jung2007rotational}
\end{aligned}
\end{equation}
where $\boldsymbol{U}$ and $\boldsymbol{\Omega}$ are the unknown lab frame linear and angular velocities of the body frame, respectively, and the overdot of $\boldsymbol \xi$ denotes a time derivative of the body frame coordinates. The above equation embodies the non-local, force-velocity relationship and non-slip boundary condition, which is augmented by the total balance forces and torques on the sperm,
\begin{equation}
\begin{aligned}
   \label{equ:2} 
   \iint_{\boldsymbol X \in \partial D} \boldsymbol{f} (\boldsymbol X) d\boldsymbol X =0 \\
   \iint_{\boldsymbol X \in \partial D} \boldsymbol X  \times \boldsymbol{f} (\boldsymbol X) d\boldsymbol X =0. 
\end{aligned}
\end{equation}
The above system of equations governs the so-called mobility problem, in which the unknown rigid-body motion results from imposed force and moment. In other words, the unknown traction $\boldsymbol{f}$, and the translational $\boldsymbol{U}$ and rotational $\boldsymbol{\Omega}$ velocities of the body frame can be obtained numerically from a prescribed waveform model relative to the body frame of reference (described below). This allows us to resolve the translating and rotating 3D kinematics of the swimming sperm at the lab frame, examples of which can be seen in Fig. \ref{image1: CompTraj ExpNum and NumModel}. The system is treated as an initial-value problem and solved via the built-in function \texttt{ode45} in \texttt{MATLAB}, and the algorithm is implemented as dimensionless, where the flagellum length is normalized to 1 and time is quantified in terms of beat cycles. Here, we consider sperm swimming in an infinite fluid and neglect boundary effects for simplicity, though this could be easily incorporated \cite{gallagher2018meshfree, ishimoto2015fluid} in future studies. The human sperm head has a marginal impact on sperm motility due to their typical small sizes (when compered against the flagellum length) \cite{ishimoto2015fluid, miki2013rheotaxis,gaffney2011mammalian}, and thus the head geometry adopted here is simplified to a scalene ellipsoid with axes of length 0.044, 0.036, and 0.022~\cite{gallagher2018meshfree}. A finer quadrature discretization with 700 points was used for the sperm head (though this could be lower), and a coarser force discretization with 136 points was used for the flagellum, with the regularization parameter $\epsilon = 0.25/45$ chosen to approximately represent the ratio of flagellar radius to length of human sperm [\textit{ibid}]. Simulations were conducted on the High Performance Computing system of University of Bristol: BlueCrystal Phase 4.

\subsubsection{Symmetric and asymmetric flagellar waveform models}

The asymmetry in flagellar beating patterns has been widely discussed in the literature \cite{brokaw1970bending, brokaw1965non, brokaw1971bend}, for example, the effect of beat plane inclination to the plane of flattening of the head in sperm boundary accumulation was investigated in~\cite{smith2009human}, though restricted to planar beatings. Transitions in swimming behaviors relevant to asymmetry and rotation were also examined recently \cite{zaferani2021rolling, zaferani2021mammalian}, however the modelling framework was limited to 2D beatings within a local hydrodynamic theory, using the so-called restive-force theory (RFT) \cite{gray1955movement}. Here we investigate the role of waveform asymmetry in 3D for freely-swimming sperm, solving the non-local hydrodynamics around sperm swimming. This allows us to reveal the intricate manifestation of asymmetry in the complex rotations and translations of sperm swimming in 3D. Below, we introduce the waveform models dictating the 3D beating patterns at the body frame coordinates $\boldsymbol \xi$, as required in Eq.~\ref{equ:1}, and provided below in Eqs. \ref{equ:3} and \ref{equ:4}. Direct experimental observations of flagellar beating relative to the sperm head so far are limited to tethered sperm \cite{ishijima1986flagellar, morgan2008tissue, khanal2021dynamic, gadadhar2021tubulin} or for 2D swimming cells \cite{gong2021reconstruction, friedrich2010high,smith2009bend}. As such, the 3D beating is approximated by an elliptical helicoidal waveform, as inferred from experiments \cite{bukatin2015bimodal, ishijima1986flagellar, powar2022unraveling, woolley2003motility} and which has long history of usage in the literature \cite{jikeli2015sperm, kantsler2014rheotaxis, smith2009human, ishimoto2015fluid, woolley2003motility}. Here we consider two static sources of waveform asymmetry that have been observed experimentally \cite{khanal2021dynamic,gadadhar2021tubulin, gong2021reconstruction, smith2009bend,friedrich2010high,zaferani2021rolling,zaferani2021mammalian}: a newly observed one-sided bias of the waveform relative to the orientation of the sperm head long axis~\cite{khanal2021dynamic}, instigated by the internal asymmetric structure of the basal body and centriole, and the asymmetric mean curvature of the flagellum over beat cycle \cite{gadadhar2021tubulin,gong2021reconstruction,smith2009bend,friedrich2010high}. Mathematically, we consider these two forms of asymmetry as follows: (a) a waveform side-shift of an otherwise symmetric waving motion, captured by the parameter $B$, and (b) a static curvature bias, $\kappa_0$, that deforms the flagellum into a static curved shape from which symmetric waving component is overlaid, see details below. The two waveform models are denoted as `$xyz$-model' and `$\kappa$-model' for simplicity, with their out-of-plane motion  regulated by the parameters $\alpha$ and $\tau$, respectively, also referred as the waveform rotation amplitude.

The $xyz$-model employs the one-sided bias beating asymmetry (Fig. \ref{image2: NumModel TraceClass ProgVel} A-F). In this case, the waveform $\boldsymbol{\xi}$-coordinates, as required in Eq. \ref{equ:1}, are prescribed directly at the body frame of reference,
\begin{equation}
\begin{aligned}
   \label{equ:3} 
  \xi_1 &= A\left[\cos(k \xi_3- t)+B\right] \\
  \xi_2 &=-\alpha \,A \sin(k \xi_3-  t), 
\end{aligned}
\end{equation}
where $A=0.2 \xi_3$ is the modulating amplitude growing linearly with $\xi_3$ \cite{ishimoto2016mechanical, ishijima1992rotational}, $k$ is the wave number, taken to be $k=\pi, 2\pi, 3\pi$ according to the estimations from the observed waving patterns \cite{mojiri2021rapid, hansen2021multifocal, khanal2021dynamic}, $B$ introduces the static one-sided shifting asymmetry, and $\alpha$ captures the out-of-plane motion of the beat, responsible for the rotation amplitude of the sperm flagellum in the body frame. If $B=0$, the waveform is perfectly symmetric (Fig. \ref{image2: NumModel TraceClass ProgVel} A and F), otherwise it yields an average flagellum shifted sideways relative to the head long axis $\boldsymbol{\xi}_3$ (Fig. \ref{image2: NumModel TraceClass ProgVel} B-E). If $\alpha=0$, the waveform is planar (Fig. \ref{image2: NumModel TraceClass ProgVel} A-B), whilst when $\alpha$ increases, the flagellum follows an elliptical path in the cross section, with perfect circular trajectories when $\alpha=1$, see the projected point clouds in Fig. \ref{image2: NumModel TraceClass ProgVel} C-F. The sign of $\alpha$ dictates the chirality of flagellar beat, with a positive sign inducing a left-handed helicoid (Movie S8) and a negative sign inducing a right-handed helicoid (Movie S9). The rotational direction of the head spinning around its longitudinal axis $\boldsymbol{\xi}_3$ is opposite to that of the tail due to the total momentum balance \cite{smith2009human}, see Movies S8 and S9. Numerical simulations using $k=2\pi$ are provided in the main text, and $k=\pi, 3\pi$ are supplied in Supplementary Information.

The second type of waveform asymmetry due to a curvature bias is introduced via the $\kappa$-model (Fig. \ref{image2: NumModel TraceClass ProgVel} I-N), in which waveform curvature $\kappa$ and torsion $\tau$ are prescribed instead,
\begin{equation}
\begin{aligned}
   \label{equ:4} 
  \kappa = \kappa_0 + A_{\kappa} \cos(k s-  t).
\end{aligned}
\end{equation}
$\kappa_0$ represents the static curvature over one beat cycle, and $A_{\kappa}$ is the amplitude. According to the experimental measurements \cite{gadadhar2021tubulin, gong2021reconstruction, jikeli2015sperm, friedrich2010high}, we take $\kappa_0$ to range from 0 to 0.6, and $A_{\kappa}$ is chosen to be 1.
If $\kappa_0=0$, the waveform is symmetric (Fig. \ref{image2: NumModel TraceClass ProgVel} I, L and M), while a non-zero $\kappa_0$ gives rise to a curved average shape of the flagellum (Fig. \ref{image2: NumModel TraceClass ProgVel} J, K and N). The out-of-plane component is controlled by $\tau$, with $\tau=0$ generating a planar waveform (Fig. \ref{image2: NumModel TraceClass ProgVel} I-J), and a larger $\tau$ producing a larger rotation amplitude of the flagellum at the body frame, with rounder waveform cross section with increasing $\tau$ (Fig. \ref{image2: NumModel TraceClass ProgVel} K-N). For comparison purpose, the wave number $k$ for the $\kappa$-model was set to $2\pi$.  
With specified curvature and torsion, the flagellum waveform is obtained by integrating the local Frenet-Serret system of equations,  
 \begin{equation}
\begin{aligned}
   \label{equ:5} 
  \frac{d \boldsymbol \xi}{ds}=\boldsymbol T,
  \frac{d \boldsymbol T}{ds}=\kappa \boldsymbol N,
  \frac{d \boldsymbol N}{ds}=-\kappa \boldsymbol T+\tau \boldsymbol B,
  \frac{d \boldsymbol B}{ds}=-\tau \boldsymbol N.
\end{aligned}
\end{equation}
where $d/ds$ is the derivative with respect to arclength, and $\boldsymbol{T}, \boldsymbol{N}, \boldsymbol{B}$ represent the tangent, normal, and binormal unit vectors of the local Frenet–Serret frame, from which the body frame coordinates $\boldsymbol \xi$ of the prescribed flagellum, used in Eq. \ref{equ:1}, is obtained.

\section{Acknowledgements.} 
The authors thank Professor Jonathan Rossiter for his inspiring and fruitful discussions. We acknowledge the computational facilities and team of the Advanced Computing Research Centre, University of Bristol: \href{http://www.bristol.ac.uk/acrc/}{http://www.bristol.ac.uk/acrc/}. Xiaomeng Ren acknowledges financial support of China Scholarship Council through Grant 202006830002.

\bibliography{MT}
%\bibliography{example}

\end{document}